\UseRawInputEncoding
\documentclass[%
 aip,
 amsmath,amssymb,
 reprint,%
]{revtex4-1}
\usepackage{graphicx}
\usepackage{color}
\usepackage{floatrow}
\usepackage{ragged2e}

\begin{document}

\title{Hybrid Stochastic Synapses Enabled by Scaled Ferroelectric Field-effect Transistors}%

\author{A N M Nafiul Islam}
\author{Arnob Saha}%
\affiliation{ 
School of Electrical Engineering \& Computer Science, The Pennsylvania State University, University Park, PA 16802, USA
}%

\author{Zhouhang Jiang}
\author{Kai Ni}
\affiliation{ 
Microsystems Engineering Ph.D. program, Rochester Institute of Technology, Rochester, NY 14623, USA
}%


\author{Abhronil Sengupta}
\email{sengupta@psu.edu}
\affiliation{ 
School of Electrical Engineering \& Computer Science, The Pennsylvania State University, University Park, PA 16802, USA
}%

\begin{abstract}
{\small
Achieving brain-like density and performance in neuromorphic computers necessitates scaling down the size of nanodevices emulating neuro-synaptic functionalities. However, scaling nanodevices results in reduction of programming resolution and emergence of stochastic non-idealities. While prior work has mainly focused on binary transitions, in this work we leverage the stochastic switching of a three-state ferroelectric field effect transistor (FeFET) to implement a long-term and short-term 2-tier stochastic synaptic memory with a single device. Experimental measurements are performed on a scaled 28nm high-$k$ metal gate technology-based device to develop a probabilistic model of the hybrid stochastic synapse. In addition to the advantage of ultra-low programming energies afforded by scaling, our hardware-algorithm co-design analysis reveals the efficacy of the 2-tier memory in comparison to binary stochastic synapses in on-chip learning tasks -- paving the way for algorithms exploiting multi-state devices with probabilistic transitions beyond deterministic ones.}
\end{abstract}

\maketitle

Interest in ferroelectric memory for storage and computing applications has been rejuvenated with the discovery of ferroelectricity in CMOS-compatible and scalable doped $\textrm{HfO}_2$, beyond conventional perovskite ferroelectrics  \cite{ferri2021ferroelectrics,boscke2011ferroelectricity,boscke2011phase}. Leveraging $\textrm{HfO}_2$, ferroelectric field effect transistors (FeFETs) are poised as top candidates for hardware tailored to future data-centric applications. The FeFET structure, shown in Fig. 1(a), mimics that of a traditional MOSFET with a ferroelectric thin film layer in the gate stack. Generally, polarization of the ferroelectric layer affects the underlying channel charge density and thus modulates the threshold voltage (\textit{V}\textsubscript{TH}) of the device. By partially switching the polarization, \textit{V}\textsubscript{TH} and consequently the channel conductance, the FeFET can be gradually tuned. This phenomenon has been used to realize multi-state nonvolatile weight cells or synaptic memory elements previously \cite{mulaosmanovic2017novel,jerry2017ferroelectric, sun2018exploiting,saha2021intrinsic}. To realize a large number of stable analog states, the size of the FeFET synaptic devices need to be very large, i.e., on the order of several $\mu m$. In addition, non-linearity in the partial switching regime requires complex pulsing schemes for compensation and achieving desired conductance changes -- adding further peripheral overhead and inefficiencies. While recent efforts have looked at leveraging the non-linearity as an advantage \cite{saha2021intrinsic}, the large area coupled with higher power requirements of the larger devices make them unattractive for large-scale integration on chip. 

To reap the benefits of great scalability afforded by Hafnia and maximize memory density, continual scaling of FeFET is desirable. However, this poses a conspicuous challenge for analog synapse applications. As these devices are scaled, the thin ferroelectric layer is unable to accommodate many domains owing to their finite size\cite{lee2021domains}. The total number of stable conductance states, which is a direct function of domain number, thus, decreases significantly. Fig. 1(b) shows this effect with the device model developed later in the article. This ultimately results in reduction of representation precision and massive degradation of performance in learning tasks. Additionally, due to the inhomogeneous distribution of coercive field of the domains and varying grain size and orientation, each domain switching event is abrupt and stochastic in nature \cite{deng2020comprehensive,mulaosmanovic2017switching}. Nonetheless, when the domain number is large, this behavior is smoothed out over the bulk and only deterministic conductance change is observed\cite{saha2021intrinsic}. When scaled down, however, the conductance change of the device itself becomes abrupt and stochastic with appreciable cycle-to-cycle variability. An exploration of scaled FeFETs and their intrinsic stochasticity is thus required.

Note, the reduction of programming resolution and emergence of stochasticity in scaled technology nodes is not unique to FeFETs but exist in many other emerging memristive technologies as well \cite{querlioz2015bioinspired,suri2013bio,vincent2015spin}. This has thus prompted exploration from different device perspectives into divergent models of computation. Motivated by the stochastic release of neurotransmitters in the biological brain regulating signal propagation \cite{branco2009probability}, the focus there has primarily been on binary stochastic synapses with probabilistic learning rules \cite{suri2013bio, srinivasan2016magnetic}. However, these systems have suffered from deteriorated learning accuracy in comparison to their higher precision memristive counterparts. To compensate, multi-device synapses have been proposed \cite{srinivasan2016magnetic, sengupta2018stochastic}. This, however, diminishes the fundamental advantage of scaling the devices in the first place. Moreover, complex probabilistic learning rules have been proposed to alleviate some performance issues, albeit at the cost of greater peripheral costs \cite{sengupta2018stochastic}.

In this work, we explore an alternative simplified learning scheme enabled by the stochastic switching characteristics of a scaled FeFET with three conductive states to facilitate efficient learning with minimal overhead. By probabilistically switching between the states, we can realize a significance driven two-tier hybrid synapse using a single device, similar to long-term (LT) and short-term (ST) memory formation in the brain \cite{atkinson1968human}. In humans, incoming sensory information are first stored in an intermediate or ST ``working" memory. Stronger repeated stimulation then results in the information moving on to the LT memory. Similarly, for our devices, correlated patterns are potentiated to an intermediate conductive state probabilistically, while further stronger correlations consolidate the patterns in a higher conductive state with a higher probability. We develop a device-circuit-algorithm co-simulation framework to assess the efficacy of such a mechanism for a large-scale spiking neural network and to ultimately implement compact, energy and area-efficient hardware for edge intelligence. Furthermore, from an algorithmic perspective, this work looks at hybrid stochastic switching beyond binary states and paves the way to explore the advantage of switching between multi-level states in a probabilistic manner rather than just deterministically in any scaled device technology.

\begin{figure}[t]
\centering
\includegraphics[width=3.41in]{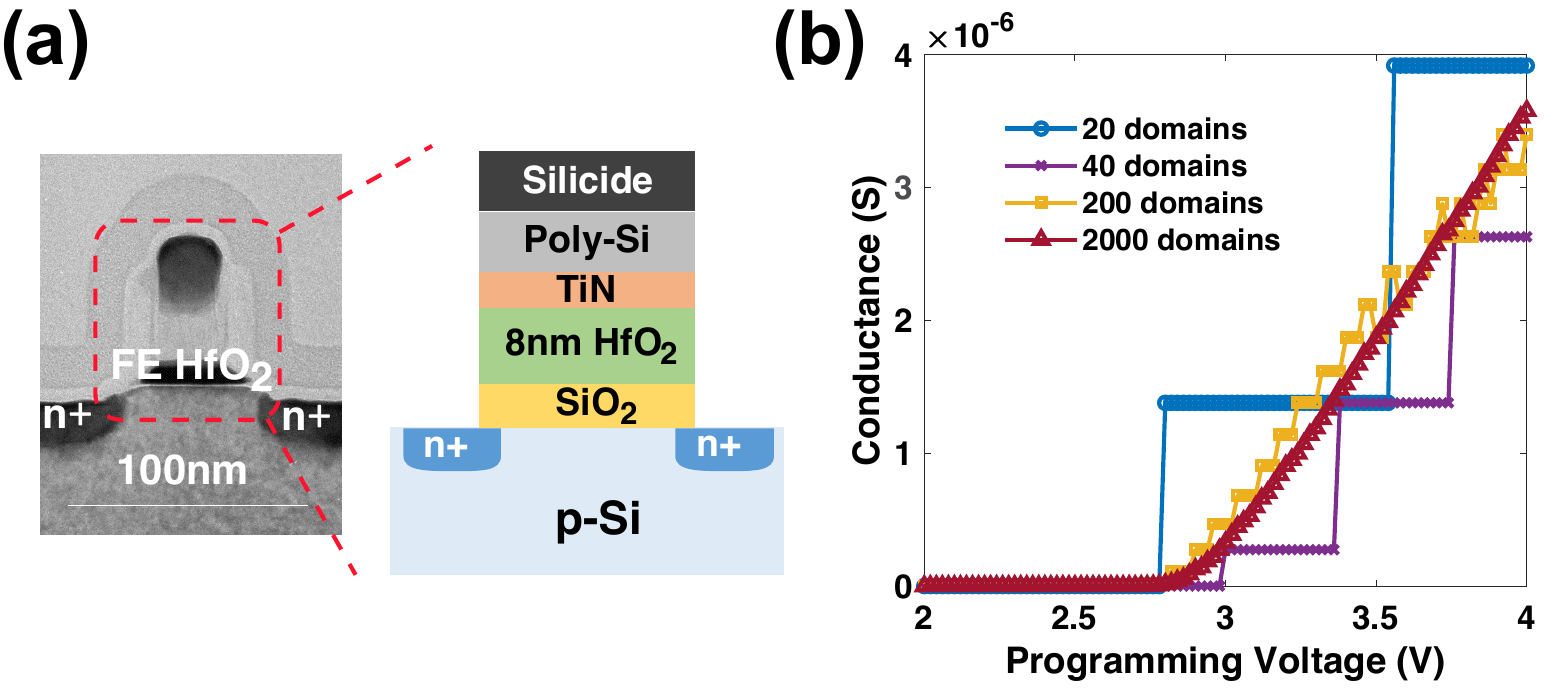}
\caption{\scriptsize{(a) TEM cross-section and schematic of the FeFET structure used in this work, with HfO\textsubscript{2} ferroelectric layer. (b) Simulation studies, using experimentally calibrated device simulation model developed in this work, showing conductance of the device with variable programming voltage magnitudes for different number of domains in the ferroelectric layer of the device. As the domain number increases, the number of available states (indicated by specific conductances) increases and their transition is no longer abrupt and stochastic.}}
\label{intro}
\end{figure}

We start by building a comprehensive device model calibrated to experimental results to understand how the states in the FeFET devices undergo stochastic switching as their dimensions are scaled. Hardware measurements were carried out for a scaled FeFET based on an industrial 28$nm$ high-$k$ metal gate (HKMG) technology \cite{trentzsch201628nm}. The device dimension is $0.24\mu m \times 0.24\mu m$ and has an $8nm$ thick doped $\textrm{HfO}_2$ ferroelectric layer on top of a thin $\textrm{SiO}_2$ oxide layer. The device is input with a pulsing scheme, shown in inset of Fig. 2(a), consisting of a reset pulse ($-4V$), and a variable programming pulse. The reset pulse resets all the domains of the ferroelectric layer to the negative polarization state. Measurements are performed for different programming voltages ranging from $2V$ to $4V$ with $20mV$ steps. For each programming voltage, the experiment is repeated 50 times to get the accumulative switching statistics. 

In Fig. 2(a), we observe the change of threshold voltage after each application of the programming voltage. We find that there are two discrete switching events taking place for this device -- one at $\sim2.9V$ and the other at $\sim3.75V$. The three states, referred henceforth as $S_0$, $S_1$, $S_2$, are clearly identifiable from the change in the threshold voltage at $\sim 1.5V$, $\sim0.5V$ and $\sim0.2V$ respectively. Fig. 2(b) shows the accumulative switching probability percentage of the device going from $S_0$ to $S_1$ and $S_0$ to $S_2$ with respect to write voltage. Switching probability of going from $S_0$ to $S_1$ increases as the programming voltage is increased. After saturating to 1, for higher programming voltages the device begins to transition more into the higher conductive state, $S_2$, and thus the probability of $S_1$ goes down with increasing voltage. The discussion holds true if we begin from the intermediate state, $S_1$, as well. For clarity, we omitted showing the probability of going from $S_1$ to $S_2$ transition. Here, $R_{S2}<R_{S1}<R_{S0}$, where $R$ denotes the resistance of the state. 

To capture the stochastic device behavior in our device-algorithm co-simulation framework, we developed a Monte Carlo Algorithm-based model\cite{alessandri2019monte,deng2020comprehensive} with the ferroelectric layer assumed to consist of multiple independent domains. The domains can be either positively or negatively polarized. A domain switch within a certain time step, $\Delta t$, is simulated with switching probability, $p_i$, by, 

\begin{equation}
\label{switching probability}
p_i (t) = 1-e^{({\frac{t}{\tau_i}})^\beta - ({\frac{t+\Delta t}{\tau_i}})^\beta}
\end{equation}

\noindent Here, $\beta$ is the shape parameter of the probability distribution, and $\tau_i$ is the switching time constant of the $i$-th domain and follows the domain-nucleation model \cite{mulaosmanovic2017switching}. Note, for constant electric field $\tau$ is constant\cite{alessandri2019monte}. 

If the $i$-th domain switches according to Eq. (1), the state of that domain is flipped. We measure the total polarization by taking the summation of all the states over the total number of domains. Note, the number of domains is a tunable model parameter. As domains switch, the internal electric field varies over time, i.e., the switching time constant becomes a function of the activation field, $E_{a,i}$, and applied electric field, $E_{fe}$\cite{alessandri2019monte}. This give the switching a history component and we capture this effect using the history parameter, $h_i (t)$, defined by,

\begin{equation}
\label{history parameter}
h_i(t) =\int_{t_0}^{t} \frac{dt'}{\tau_i(E_{fe}(t'),E_{a,i})}
\end{equation}
 
When starting from complete reset, the history parameter, $h_i (t)$, increases temporally until all the domains are inverted. The domain switching probability, $p_i$, thus can be rewritten as,

\begin{equation}
p_i(t) = 1-e^{(h_i(t))^\beta - (h_i(t+\Delta t))^\beta}
\end{equation}

\begin{table}[!b]
\label{table1}
\center
\centerline{Table I: Device Model Parameters}
\vspace{2mm}
\begin{tabular}{c c}
\hline 
\bfseries Parameters & \bfseries Value\\
\hline
Number of domains, $N_{dom}$ & $20$\\
Time-step, $\Delta t$ & $20ns$\\
Constant polarization for each domain & $25\mu C/cm^2$  \\
Polarization time constant, $\tau_0$ & $1.9\times10^{-8}s$\\
GB2 Parameters, $(a,b,p,q)$ & $(2.45,0.2,0.6775,0.8115)$\\
Shape parameter, $\beta$ & $2$\\
Temperature & $300K$\\
Substrate Doping & $3\times 10^{17}cm^{-3}$\\
$\textrm{SiO}_2$ interlayer thickness & $1nm$ \\
$\textrm{HfO}_2$ thickness & $8nm$ \\
Read gate voltage & $1.2 V$\\
Read drain voltage & $0.05V$\\
Read source voltage & $0 V$\\
\hline 
\end{tabular}\\ 
\end{table}

\begin{figure}[t]
\centering
\includegraphics[width=3.41in]{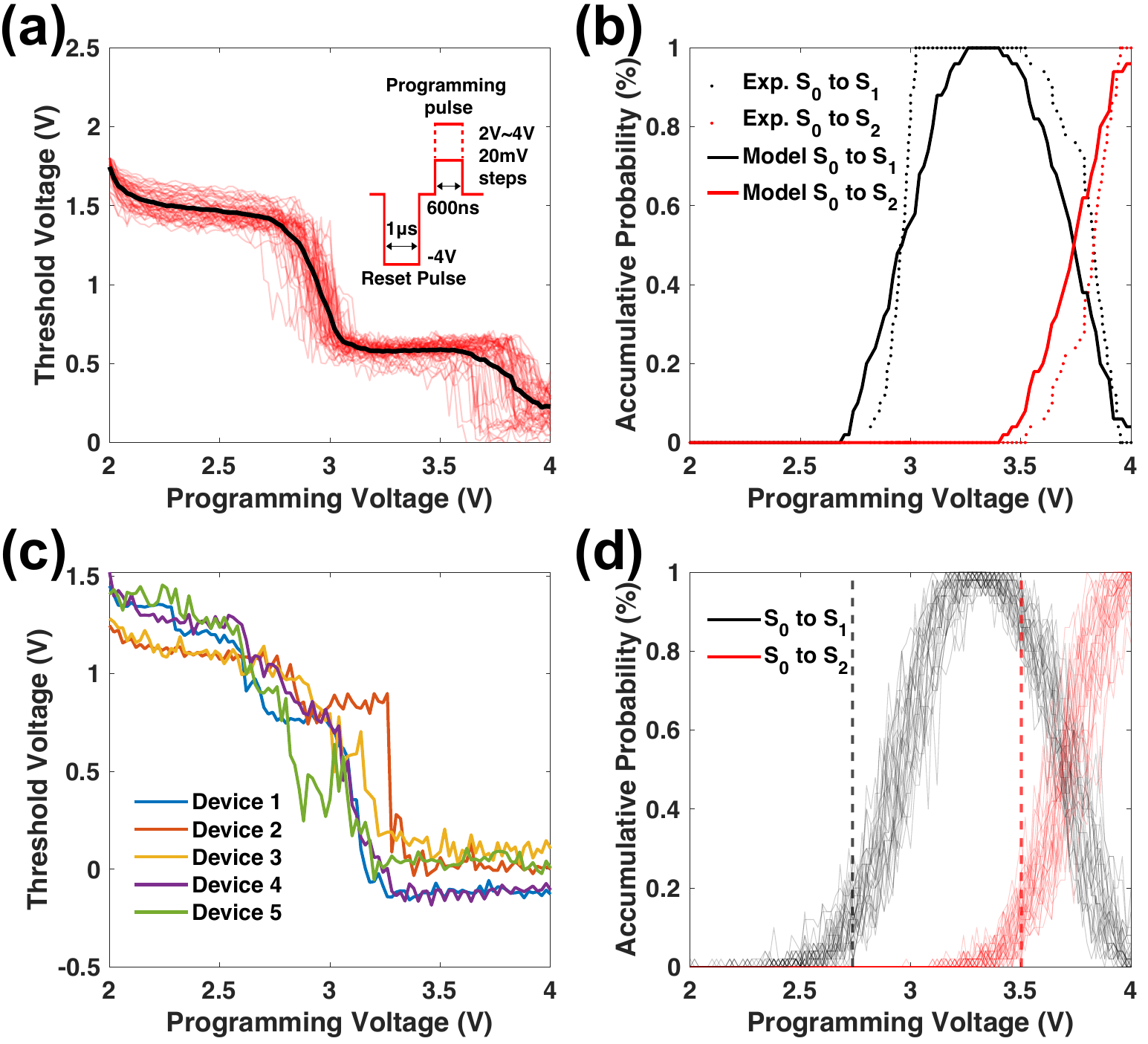}
\caption{\scriptsize{(a) Variability of threshold voltage with programming voltage amplitude over 50 iterations. The black line represents the average of the 50 iterations (red). The inset shows the pulsing scheme for potentiating the device for extraction of threshold voltage consisting of a reset pulse of magnitude = $-4V$ and a programming pulse of magnitude = $2V$ to $4V$ with $20mV$ increments. (b) Accumulative switching probability of the device from $S_0$ to $S_1$ and from $S_0$ to $S_2$ for the 50 repetitions with experimental and model data. (c) Threshold voltage with respect to programming voltage for 5 individual devices over a single iteration. The devices show similar operation ranges. (d) Simulation result of accumulative switching probability of 30 different devices with applied programming voltage over 50 runs. We observe a range of switching probabilities for any given voltage as indicated by the vertical lines. Black lines represent switching from $S_0$ to $S_1$ and red lines indicate switching from $S_0$ to $S_2$.}}
\label{exp}
\end{figure}

Along with the history parameter, the model is self-consistently solved for charge-voltage equations to obtain the final FeFET device characteristics. The device model parameters used for the simulation can be found in Table I. We fit the accumulative transition probabilities from $S_0$ to $S_1$ and from $S_0$ to $S_2$ with our device model and find that by varying the number of domains in the ferroelectric layer we can capture the effect of scaling. For a domain number of 20, we obtain the closest fit to the experimental probabilities (Fig. 2(b)).

To address device-to-device variability, we characterize additional devices using the previously discussed measurement scheme. Fig. 2(c) shows the threshold switching dynamics of 5 individual devices. The devices display similar switching voltage ranges. Note, this variation in switching is already captured by our model using the electric field, $E_{fe}$, whose probability density function follows a generalized beta distribution of type 2 (GB2), \cite{alessandri2019monte}. In Fig. 2(d), we simulate our model for 30 different devices with the same pulsing scheme as before for 50 iterations each and observe that for a single applied programming voltage (indicated by vertical lines), we can get a range of switching probabilities (upto $\sim20\%$) across the various devices. For a system-level implementation of these devices, the inter-device randomness needs to be considered.

\begin{figure}[t]
\centering
\includegraphics[width=3.41in]{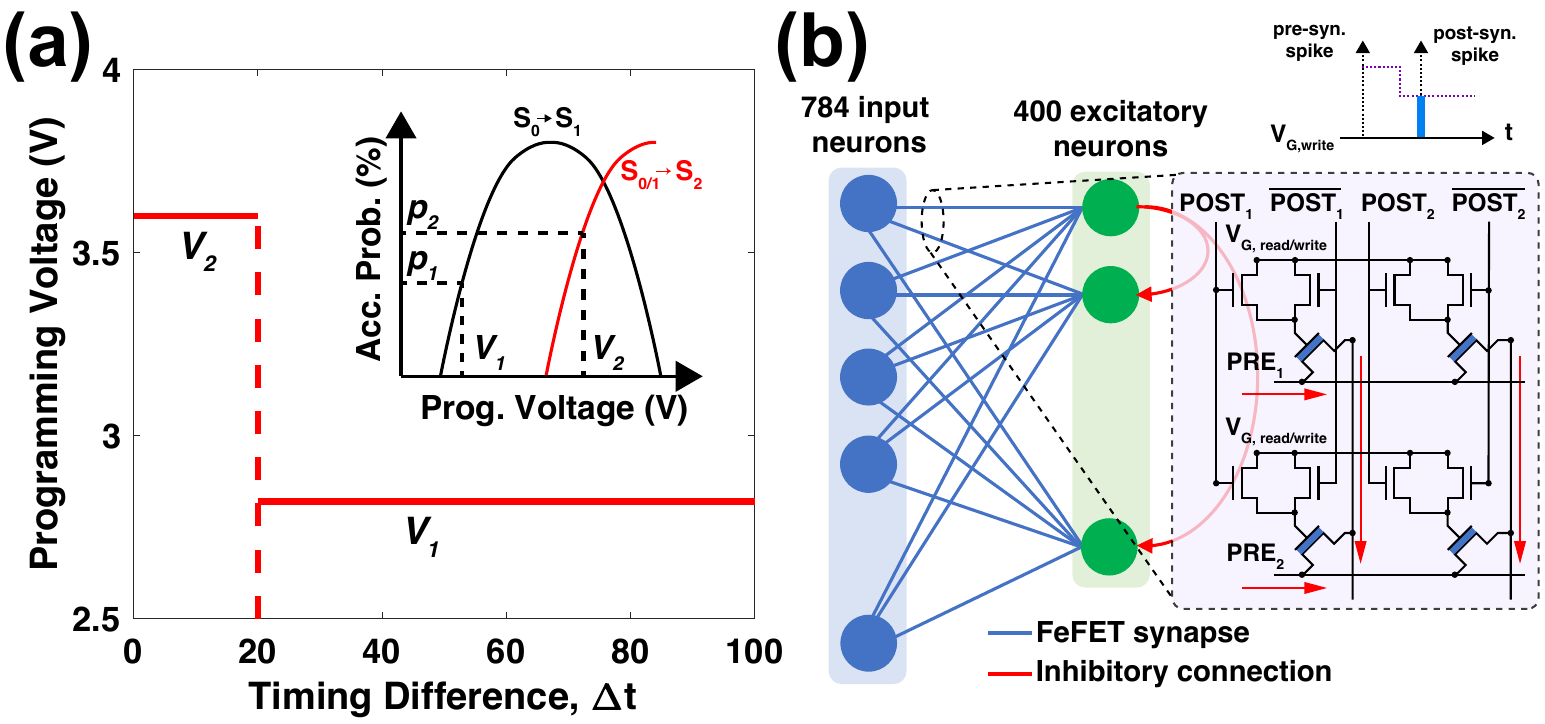}
\caption{\scriptsize{(a) Learning rule for the Spiking Neural Network. When the timing difference between the pre- and the post-synaptic spike is small, the network is potentiated towards the higher conductive state with a probability, $p_2$, by application of voltage pulse, $V_2$, while for greater differences, the synapse is potentiated to the intermediate state with probability, $p_1$, by voltage pulse, $V_1$. Here, $p_2>p_1$. The inset cartoon shows the two operating points on the accumulative switching probability curve. (b) The network architecture for training. It can be embedded in a cross array, shown in inset, where the input spikes coming from PRE are modulated by the FeFET conductances and are summed up along the columns. V\textsubscript{G, read/write} provides the necessary gate voltage to read or write the device as necessary. Access transistors are controlled based on the timing of pre- and post-synaptic spikes to ensure proper operation.}}
\label{learning}
\end{figure}

The stochastic transition from a low conductance state to an intermediate state to finally a high conductance state opens up opportunities to explore in neuromorphic algorithm design. In the human brain, memories are usually first stored in ST memory. Stronger stimulation then urges the brain to consolidate that into LT memory. Our synaptic device dynamics is analogous to this memory formation. We use our calibrated device model to explore how such a 2-tier memory, which can be thought of as a hybrid memory combining LT ST memory, can help in learning in comparison to traditional binary stochastic synapses in neuromorphic systems -- specifically for Spiking Neural Networks (SNNs) \cite{sengupta2019going}. Akin to the brain, electrical pulses (referred as ``spikes") are propagated through SNNs to perform necessary computation. Besides bio-plausibility, SNNs are energy-efficient and hardware-friendly for on-chip intelligence, especially in resource constrained scenarios\cite{aimone2021roadmap}.
 
As an unsupervised training alternative, SNNs are trained using local, unsupervised rules, such as Spike-timing dependent plasticity (STDP) \cite{bi1998synaptic}, where the timing difference between spiking events of the pre-synaptic and post-synaptic neurons are used to modify the synaptic weights, usually in an exponential manner \cite{sengupta2016hybrid}. However, since our FeFET synapses have only 3 states with stochastic switching in-between, the learning approach requires rethinking. We draw inspiration from the human LT ST memory and adopt a stepped learning rule, given by Fig. 3(a). When a pre-synaptic spike is closely followed by a post-synaptic spike indicating a strong temporal correlation, the weights are potentiated to $S_2$ with a higher probability, $p_2$, by applying voltage pulse of amplitude $V_2$. Similarly, when the timing difference is greater than a certain threshold, signaling a comparatively weaker correlation, we potentiate to state $S_1$ with smaller probability $p_1$ by applying a lower voltage pulse, $V_1$. In this scenario, if the synapse is already in $S_1$, the state remains unchanged. Note, the probabilistic nature of these transitions allow the network to generalize during training. Additionally, the tunability of the accumulative probability ($p_1, p_2$) with programming voltage allows us great flexibility in designing the network. Such a learning rule with a 2-level voltage pulse ($V_2=3.6V$ and $V_1=2.82V$ in our case) can be input directly to a cross array of devices for programming, thus making the hardware overhead minimal as shown in Fig. 3(b) inset. Access transistors are required for each of the synaptic devices to decouple the ``read" and ``write" paths for online learning. During programming, after a pre-synaptic spike the 2-level pulse is applied to $V_{G,write}$. This is sampled to program the synapse appropriately whenever there is a post-synaptic spike, which activates the corresponding POST signal. During the read operation, input spikes coming from PRE are modulated by the synaptic conductances and are summed up along the columns to feed into the neurons. Additionally, we find from prior works that potentiation plays a far significant role than depression in such probabilistic learning rules \cite{srinivasan2016magnetic,koo2020sbsnn,srinivasan2019restocnet}. Thus, we decide to forego depression in the algorithm altogether, further simplifying the circuit overhead required for programming.

\begin{table}[!b]
\label{table1}
\center
\centerline{Table II: Network Simulation Parameters}
\vspace{2mm}
\begin{tabular}{c c}
\hline 
\bfseries Parameters & \bfseries Value\\
\hline
Number of neurons & $400$  \\
Batch-size & $1$\\
Neuron threshold voltage, $\theta_0$ & $-52mV$ \\
Resting potential, $v_{rest}$ & $-65mV$ \\
Membrane reset potential, $v_{reset}$ & $-60mV$ \\
Refractory period, $\delta_{ref}$ & $5ms$ \\
Time constant of neuron voltage decay, $\tau_{neuron}$ & $100ms$ \\
Adaptive threshold voltage increment, $\theta_+$ & $10.0$ \\
Static inhibitory synaptic weight, $w_{inh}$ & $-480$ \\
Maximum weight & 3.0 \\
Maximum firing rate & $128$\\
Timing difference threshold, $T_{diff}$ & 20ms\\
\hline 
\end{tabular}\\ 
\end{table}

We evaluate our device framework and the algorithm enabled by its unique capabilities for learning in a network setting with the scaled FeFETs as the synaptic weights. The network was trained on the MNIST handwritten digit recognition dataset\cite{lecun1998mnist} using a modified PyTorch-based package, BindsNET\cite{hazan2018bindsnet}. The network architecture\cite{diehl2015unsupervised}, shown in Fig. 3(b), consists of 784 input neurons followed by 400 excitatory leaky-integrate-fire (LIF) neurons\cite{ghosh2009spiking} connected through the 2-tier synapses. The LIF neurons' membrane voltage, $V_{mem}$ is governed by,

\begin{equation}
\label{LIF equation}
\tau_{neuron} \frac{dV_{mem}}{dt} = -(V_{mem}-v_{rest})+I_{input}
\end{equation}

\noindent Here, $v_{rest}$ is the neuron resting potential, and $I_{input}$ is the total input current to the neuron. When $V_{mem}$ reaches the threshold voltage ($\theta_0$), the neuron spikes and $V_{mem}$ is reset to $v_{reset}$. Additionally, each neuron undergoes a refractory period ($\delta_{ref}$), where it cannot fire again. 

The excitatory neurons are connected recurrently to all except themselves with static inhibitory connections ($w_{inh}$) ensuring the winner-take-all mechanism. The neurons implement homeostasis through adaptively increasing their threshold ($\theta_+$) such that no single neuron dominates during training. The images were converted to Poisson spike trains based on their analog pixel intensities before being input to the network. We trained the network over 1900 training patterns with five random initializations, and obtained an average accuracy of 80.70\% over the test set, with a maximum of 81.28\% (outperforming state-of-the-art iso-neuron implementations of stochastic unsupervised learning \cite{srinivasan2016magnetic,koo2020sbsnn,srinivasan2019restocnet}). Additionally, to capture the device-to-device variations, the probability of switching of the FeFET synapses was varied randomly by 20\% (as a worst case scenario, as shown in Fig. 2(d)). In this scenario, the network was able to attain an accuracy of 80.45\%, illustrating the resiliency of the on-chip stochastic learning scheme. Note that the accuracies can be increased further by increasing the number of excitatory neurons in the network. The learnt patterns are shown in Fig. 4(a) and the network simulation parameters are listed in Table II. For in depth discussion on SNNs, readers are referred to \cite{diehl2015unsupervised}. 

We compare the performance of our 2-tier FeFET synapses with binary synapses. For both cases, the same learning rule with constant probabilities is used. Since the binary state has only two levels, the device was potentiated to the high state at a high probability for highly correlated inputs while for weakly correlated spikes the device is potentiated to the higher conductance with a lower probability. Note, the timing difference threshold ($T_{diff}$) beyond which we considered the inputs to be weakly correlated is a hyperparameter of the learning rule. We observe that the completely binary system using this scheme only reaches an accuracy of 70.78\%, i.e., our scaled FeFETs are able to give us an improvement of $\sim$10\% over the binary case while being comparative in area and energy costs. Additionally, our accuracy results are within 6\% of multi-state FeFET implementations \cite{saha2021intrinsic} while being only a fraction in device footprint and power consumption, showcasing the efficacy of our hardware-software co-design platform.

We note that in actual hardware implementation of such devices in crossbar arrays, due to process variations, there might be devices which may not have 3 states available, but instead have binary states or greater than 3 states. Fig. 4(c,d) shows these two cases for 2 of our FeFET devices. Although domain engineering is necessary to give us control over the number of states and the process\cite{lee2021domains}, it is worthwhile to consider the effect of fewer or more states on the network implementation. For greater than 3 states, our algorithm holds true without loss of generality, as one of the intermediate states can be thought of as the short-term memory state. We found that the network can tolerate variation in the conductance value of the intermediate state (for a $10\%$ variation, we observe $<1\%$ change in average accuracy over 5 runs). For the binary case, however, the network does lose fidelity. To quantify this, we ran the network simulations with a certain percentage of devices having only binary transitions. We observe a linear decrease in performance as the \% of weights with only two states is increased (Fig. 4(b)) and gradually reaching the reported accuracy on binary synapses. This further illustrates directly the capability of the LT ST memory in a single device in contrast to binary synapses. 

\begin{figure}[t]
\centering
\includegraphics[width=3.41in]{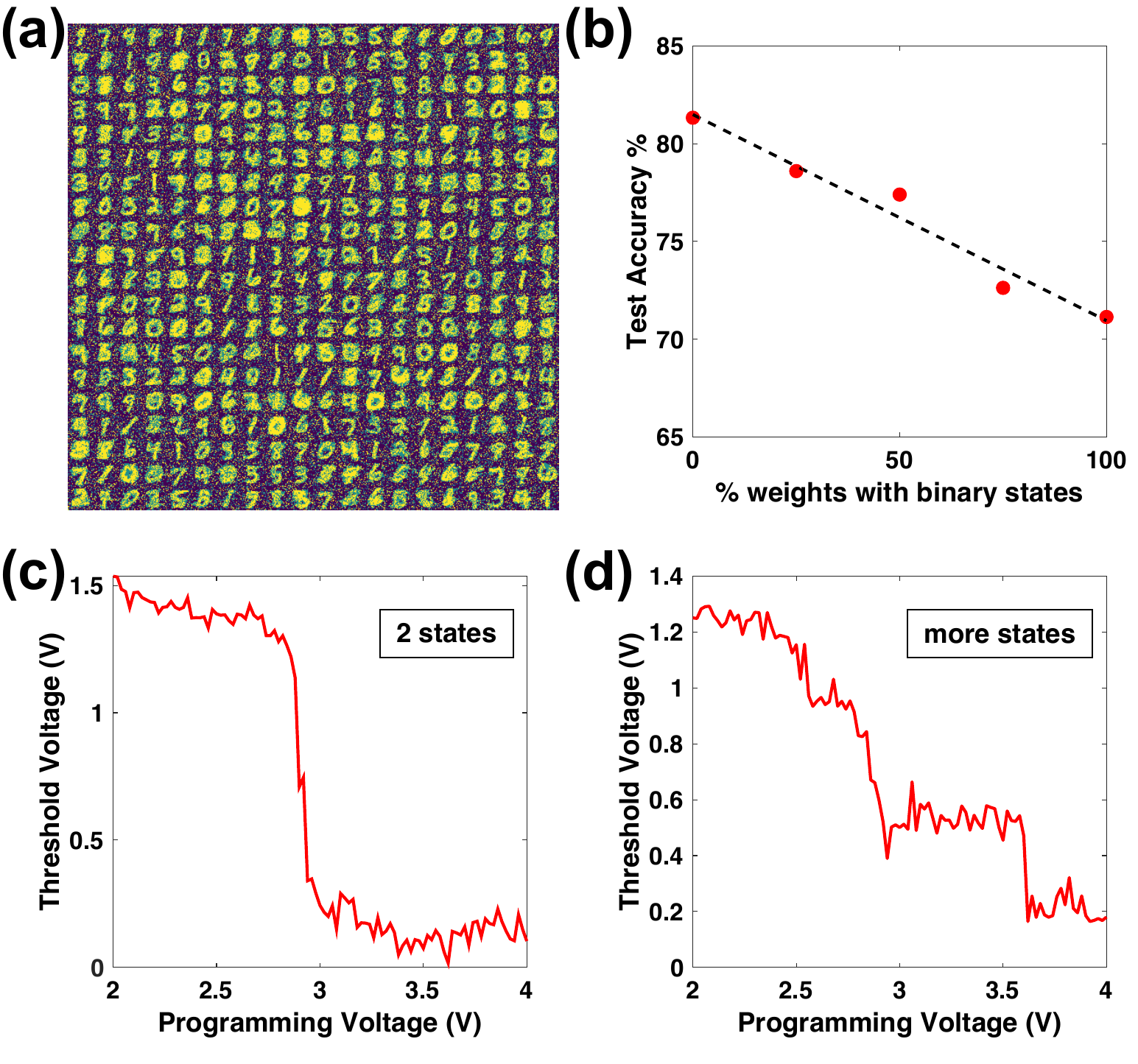}
\caption{\scriptsize{(a) Network weight patterns after training on MNIST dataset. (b) Test accuracy for different percentage of weights with two or binary states with the rest having three states. We observe a smooth degradation of performance as the \% weights with the 2-tier memory decreases. (c) A scaled FeFET device showing binary switching with change of programming voltage. (d) A scaled FeFET with more than three states (4), as indicated by the 4 distinct stable threshold voltage ranges with change of programming voltage.}}
\label{network}
\end{figure}

To summarize, we show that scaled ferroelectric field effect transistors open up an exciting direction in algorithm design with LT and ST memory in a single device. Our results are comparable to larger devices with multi-state capabilities while offering great gains over binary approaches. This not only ensures major energy and valuable area savings but also greatly expands the capabilities and efficacies of these scaled device technologies for on-chip learning applications.

\vspace{-10mm}

\section*{}
The authors would like to acknowledge GlobalFoundries Dresden Germany for providing FeFET testing devices. This material is based upon work supported primarily by the U.S. Department of Energy, Office of Science, Office of Basic Energy Sciences Energy Frontier Research Centers program under Award Number DE-SC0021118. The electrical characterization is also partially supported by SRC through GRC LMD program under task 2999.

\section*{Data Availability Statement}
The data that support the findings of this study are available from the corresponding authors upon reasonable request.

\section*{References}
%

\end{document}